\pdfoutput=1
\documentclass[final,5p,times,twocolumn]{elsarticle}
\usepackage{graphicx} 
\usepackage{epstopdf} 

\usepackage{lineno}
\begin{document}
\begin{frontmatter}
\title{Beam test results  of 3D fine-grained scintillator detector prototype for  a T2K ND280 neutrino active target}

\author[inr]{O.~Mineev\corref{cor1}}
\ead{oleg@inr.ru}
\author[ug]{A.~Blondel}
\author[ug]{Y.~Favre}
\author[inr]{S.~Fedotov}
\author[inr]{A.~Khotjantsev}
\author[ug]{A.~Korzenev}
\author[inr,mft,mif]{Yu.~Kudenko}
\author[inr]{A.~Mefodiev}
\author[ug]{E.~Noah}
\author[ug,cern]{D.~Sgalaberna}
\author[inr]{S.~Suvorov}
%\author[]{.....}

\cortext[cor1]{Corresponding author}
\address[inr]{Institute for Nuclear Research of RAS, Moscow, Russia}
\address[ug]{University of Geneva, Geneva, Switzerland}
\address[mft]{National Research Nuclear University MEPhI, Moscow, Russia}
\address[mif]{Moscow Institute of Physics and Technology, Moscow region,  Russia}
\address[cern]{European Organization for Nuclear Research (CERN), Geneva, Switzerland}

\begin{abstract}
An upgrade of the long baseline neutrino  experiment T2K near detector ND280  is currently being developed with the goal  to reduce  systematic uncertainties in the prediction of number of events at the far detector Super-Kamiokande.  The upgrade program includes the design and construction of a new highly granular fully active scintillator  detector with 3D WLS fiber readout as a neutrino target. The detector of  about $200\times 180\times 60~cm^3$ in size and a mass of  $\sim$2.2~tons will be assembled  from about $2\times10^6$ plastic scintillator cubes  of $1\times1\times1~cm^3$. 
Each cube is read out by three orthogonal Kuraray Y11 Wave Length Shifting (WLS) fibers threaded through the detector. A detector prototype made of 125 cubes was  assembled and tested  in a  charged particle test beam at CERN in the fall of 2017. This paper presents the results obtained on the light yield and timing as well as  on the optical cross-talk between the cubes. 
\end{abstract}

\end{frontmatter}
%________________________________________________________________

\section{Introduction}

The  long baseline neutrino  experiment T2K \cite{t2k}  taking data at the Japan Proton Accelerator Research Complex (J-PARC)  has obtained  a first hint on CP violation in neutrino oscillations and  excluded CP conservation at a level of significance of 2 standard deviations \cite{2sigma}.
The T2K collaboration  expects to collect higher statistics, thanks to planned upgrades to the J-PARC and the neutrino beamline, and reach a sensitivity of more than 3$\sigma$ to CP violation observation.  An upgrade of the T2K near detector  ND280 for the run extension is currently being designed with the goal to reduce systematic uncertainties in the prediction of number of events at the far detector Super-Kamiokande.  

The upgrade program of ND280  includes the development of a new highly granular fully active scintillator  neutrino detector as a neutrino target. 
The ND280  fine-grained detectors (FGD) were designed as arrays of scintillator bars disposed perpendicular to the beam axis. Geometry was optimized to detect particles propagating in the forward direction, resulting in a dependence of acceptance and resolution on the lepton scattering angle for the neutrino events.
It is also important  to study the effect of nuclear activity by measuring the energy deposited by additional nucleons originating from the interaction or resulting from nuclear breakup. In this context the natural granularity scale must be around 1~cm, corresponding to the range in plastic of protons with momentum commensurate with the Fermi motion of about 220~MeV/c.

\begin{figure}[t]
\begin{centering}
\includegraphics[width=8cm]{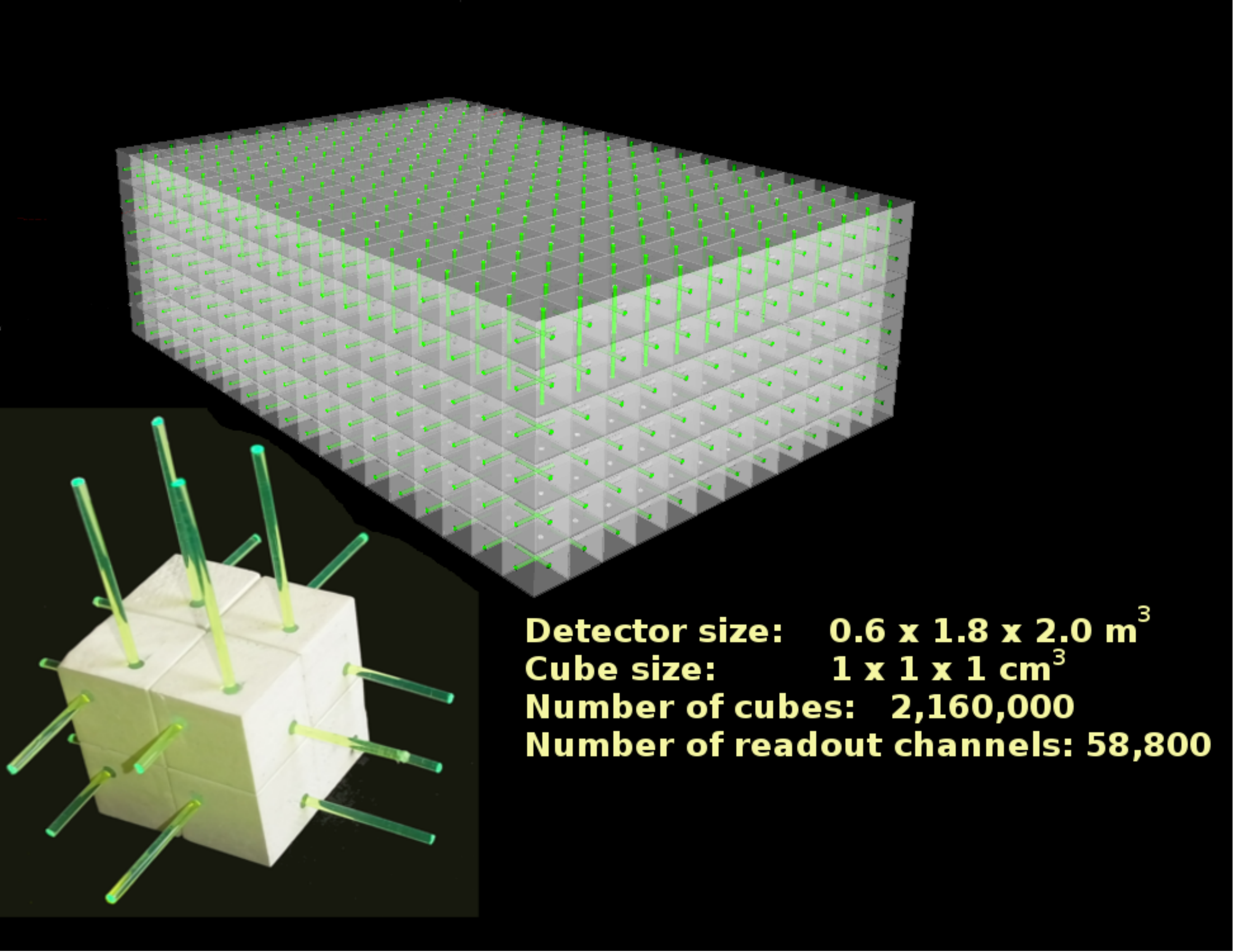}
\caption{Schematic view of the superFGD conceptual design. An array of 2x2x2 scintillator cubes illustrates the 3D readout by Wave Length Shifting (WLS) fibers.}
\label{fig:superFGD} 
\end{centering}
\end{figure}

To obtain a more isotropic response and improve the localization of the neutrino interaction vertex,  a new scintillator detector  with 3D fiber readout  was proposed as a neutrino target. The basic concept and design of this novel  highly granular fully active  detector (superFGD) are described in \cite{superFGD}. The SuperFGD concept is illustrated in  Fig.~\ref{fig:superFGD}. 
%========================
%======================== 

The SuperFGD for T2K is foreseen with dimensions of  $\sim 200\times 180\times 60~cm^3$ and a total mass of about 2.2~tons. The detector will be assembled of more than  2~millions  small cubes  of $1\times 1\times 1~cm^3$.
Each cube   has three orthogonal through holes of  1.5~mm diameter. The signal readout  from each cube is provided by   three 1.0~mm Kuraray Y11 multiclad  WLS fibers \cite{wls} connected at one end to  micro-pixel avalanche  Hamamatsu MPPC photodiodes. A single fiber reads out a whole line, row or column of cubes so that the total number of readout channels is 58800. By using the time and pulse height information from the fired fibers in all $x-y-z$ directions, the real hit cubes can be determined with little ambiguity for neutrino events in the GeV energy region, thus providing the 3D track  coordinates with an accuracy given by the cube size.   

The structure of the superFGD allows to make small prototypes to determine the operational performance  of the actual detector. A small prototype of superFGD comprised of 125 cubes  was tested in a beam of charged particles at CERN in October, 2017. Light yield, cross-talk, and time resolution are presented in this paper.
%_________________________________________________________________________

\section{Prototype design and experimental setup}

More than 10k cubes were produced at UNIPLAST Co. (Vladimir, Russia) to assemble mock-ups and prototypes. At the initial stage of R\&D the cubes were cut in size $1\times 1\times 1~cm^3$ out of long 1~cm thick extruded slabs.  For the real detector we plan to use the another production method of cubes, injection molding, which is  now under development.  Both methods provide the same light yield, the main differences are in manufacturing cost of large quantities of cubes, and in the reproducibility of geometrical size.
The scintillator composition is  polystyrene doped with 1.5\% of paraterphenyl (PTP) and 0.01\% of POPOP. After fabrication the cubes are covered by a reflecting layer by etching the scintillator surface in a chemical agent. The etching results in the formation of a white polystyrene micropore deposit over the scintillator~\cite{nim_etching}.  The thickness of the reflector layer is within 50--80~$\mu$m.  Three orthogonal through holes of 1.5~mm diameter were drilled in the cubes to accommodate  WLS fibers as shown in Fig.~\ref{fig:superFGD}. 

\begin{figure}[h]
\begin{centering}
\includegraphics[width=9cm]{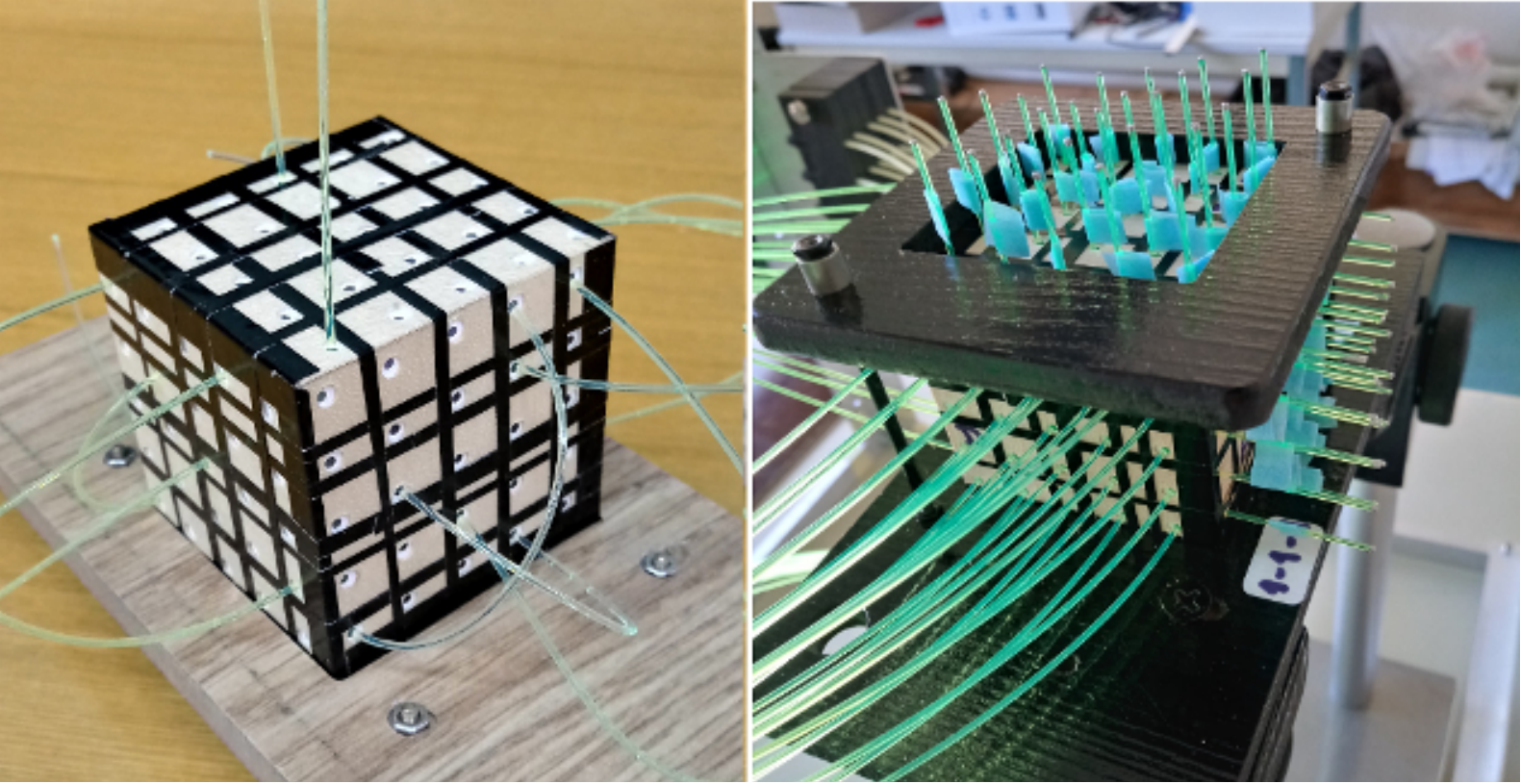}
\caption{The array of 125 cubes with 3D fiber readout. Right photo shows the array with 75 fibers mounted for the beam test. }
\label{fig:cube} 
\end{centering}
\end{figure}

A $5\times 5\times 5$ array compised of 125 cubes is  shown in Fig.~\ref{fig:cube}.  No additional layers between cubes were used, only outer protective skin made of thin hard material and adhesive tape.  
%========================
%======================== 
75 WLS fibers were inserted through the cubes, protruding 3--4~cm out of the scintillator. The fibers are 1~mm diameter  Y11(200) Kuraray S-type of 1.3~m length. One end of the fiber is attached to a photosensor, another end is covered by a reflective Al-based paint (Silvershine).
The photosensors are Hamamatsu MPPCs 12571-025C with a $1\times1~mm^2$ active area and 1600 pixels. 

We applied sequentially in the beam test two types of front-end electronics. All channels were read out by multi-channel boards developed on the basis of the OMEGA "CITIROC" ASIC for the Baby MIND spectrometer~\cite{babyMIND}. These electronics boards provide a no-dead-time readout of the detector in a 2.5 ns time slots, allowing convenient selection and reconstruction of  tracks. In order to measure the main parameters of the prototype with a high time resolution, we used custom made preamplifiers and the 16-channel CAEN digitizer DT5742 with 5~GHz sampling rate and 12-bit resolution. All results reported in this paper are hereafter obtained with the digitizer information.

\begin{figure}[h]
\begin{centering}
\includegraphics[width=8cm]{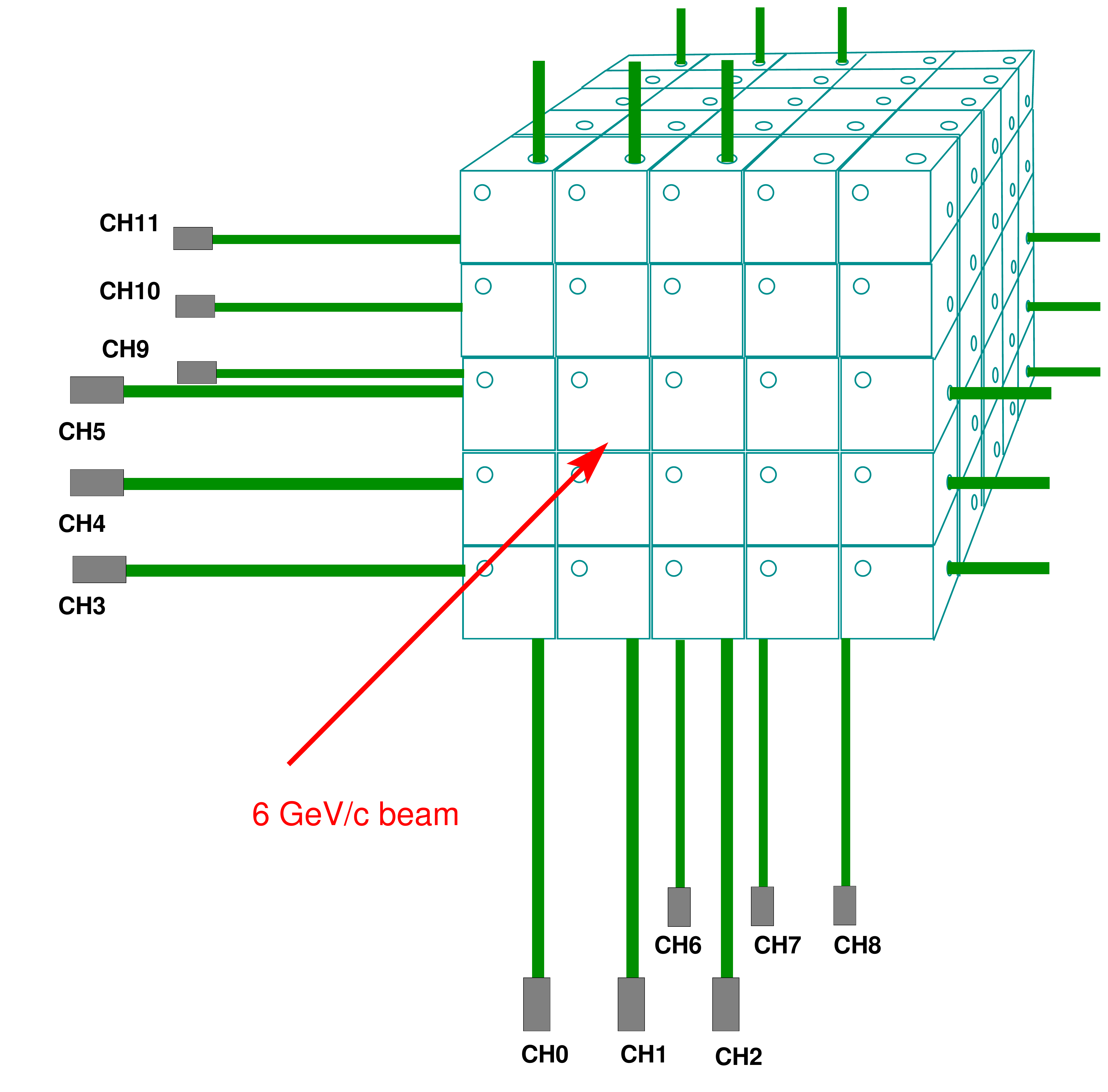}
\caption{ Readout of WLS fibers by the digitizer, and channel labeling. Inactive fibers are not shown.}
\label{fig:readout} 
\end{centering}
\end{figure}

Two  small  scintillator trigger counters of $3\times 3\times 10~mm^3$ size spaced at distance of 26~cm were installed before and after the prototype. Thus we were able to select minimum ionizing particles (MIPs) from the beam with the  position accuracy of about 3~mm. 
One of the trigger counters sent a signal to start the digitizer, signals from another one were measured by the digitizer and used in off-line analysis. Also an anti-coincidence (AC) scintillator counter with $10\times 10~cm^2$ area and a 9~mm aperture for the beam entrance was mounted in front of the prototype to remove accidentals. Trigger and AC counters were read out by the same MPPCs  12571-025C. All MPPCs were selected to have close values of the bias voltage, so we were able to fix it to 67.5~V recommended by Hamamatsu in the specification. 
In total, the digitizer reads out 12 WLS fibers, as shown in Fig.~\ref{fig:readout}, a small trigger counter and two channels from the AC counter.  The layout of readout fibers allow us to measure the parameters of 9 cubes in the front layer and 9 cubes in the back layer of the prototype. All other fibers were in place but idle for analysis.
%========================
%======================== 

The test beam was held at T10  area of the CERN Proto-Synchrotron (PS).  The line transported 6~GeV/c positive  particles  of mixed composition (mainly positrons and protons) with a momentum resolution of $\sim$0.5\%. A trigger rate of around 100~Hz has been set by closing the beam collimators in order to maximize the fraction of single-hit events.
%_________________________________________________________________________

\section{Performance in the beam}

\subsection{Light yield}

\begin{figure}[h]
\begin{centering}
\includegraphics[width=6cm]{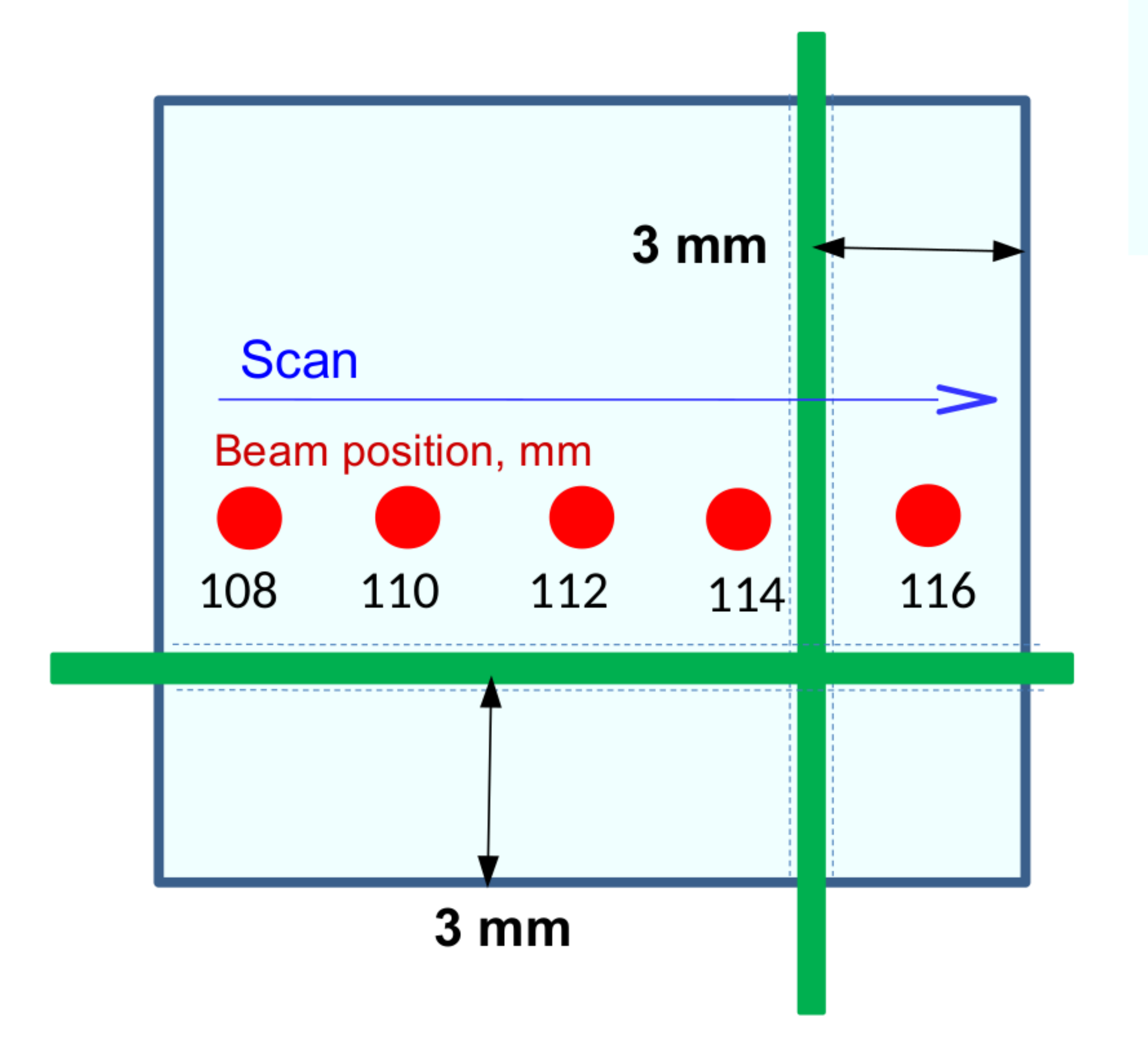}
\caption{Beam scan across a single cube. Fiber positions (in green) are shown relative to the beam hit points (in red).}
\label{fig:scancube} 
\end{centering}
\end{figure}

A beam scan with a step of 2~mm was done across 3 cubes in the horizontal direction.  Fig.~\ref{fig:scancube} shows the position of the beam center for the scan points, with respect to the position of vertical and horizontal fibers in a cube. Beam particles were localized by the trigger counters within the spot of about $3\times 3~mm^2$. The events were selected if the light yield in a small trigger counter was larger than 50~p.e. and the time difference between both trigger counters did not exceed 1~ns.

\begin{figure}[h]
\begin{centering}
\includegraphics[width=9cm]{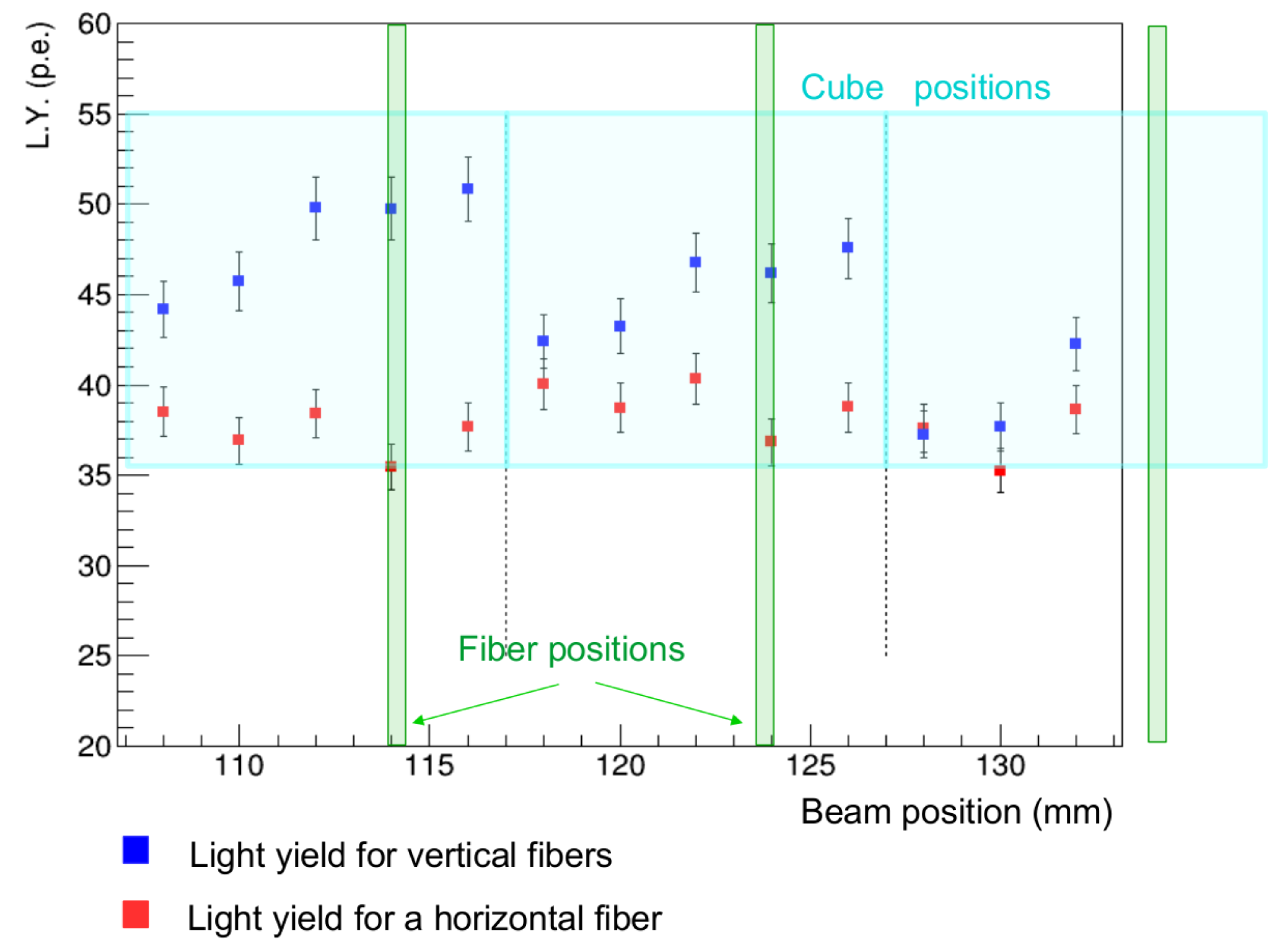}
\caption{Light yields for  horizontal and vertical fibers vs beam position. Also vertical fibers  and cube coordinates  are shown at the horizontal scale. }
\label{fig:scan} 
\end{centering}
\end{figure}

We have measured 13 scan points over a span of 25~mm, 3 cubes were scanned in the front layer along with 3 cubes in the back layer.  Each cube was read out by two fibers to measure the response as a function of the beam position. The signal charge was calculated as the area of signal waveform normalized to the signal obtained for a single photoelectron (p.e.). Calibration coefficients were calculated for each run thanks to the excellent single photoelectron response of MPPCs. The result of the scan for the front layer is presented in Fig.~\ref{fig:scan}. 
%========================
%======================== 
%========================
%======================== 

Edge effects at the cube boundaries were minimized by selecting events with  a light output exceeding the average crosstalk in both vertical and horizontal fibers. Although the beam spot is comparable to the cube size, we can observe a systematic increase in light yield when the beam point gets closer to the vertical fiber.  The horizontal fiber demonstrates fluctuations of the light signal within measurement accuracy.

The light yield (L.Y.) for different channels varies from 36 to 50~p.e./MIP for a single fiber.  The typical L.Y. was close to 40~p.e./MIP/fiber, and the total L.Y. from two fibers in the same cube was measured on an event-by-event basis to be about 80~p.e., as expected.

\subsection{Optical crosstalk}

Since  the white chemical reflector, like any reflector of the diffuse type, does not fully isolate the scintillation light, the leakage of light from a fired cube to the neighboring ones was investigated. Crosstalk was measured on an event-to-event basis as the ratio of signals in adjacent cubes to the signal in the fired cube. The average of the distribution of these ratios was defined as the average crosstalk. 
Accidentals and induced electronic noise increase the pedestal fluctuations and create a false crosstalk. To suppress this contribution we considered the signal less than 0.5 p.e. as a zero value (pedestal).
The dark noise of the MPPCs generates accidental single p.e. signals.  We have measured that the dark noise adds  less than ~0.2\% to the total value of crosstalk, thanks to the low level of dark rate of MPPCs S12571-025 ($\sim$100~kHz typical value).

\begin{figure}[h]
\begin{centering}
\includegraphics[width=9cm]{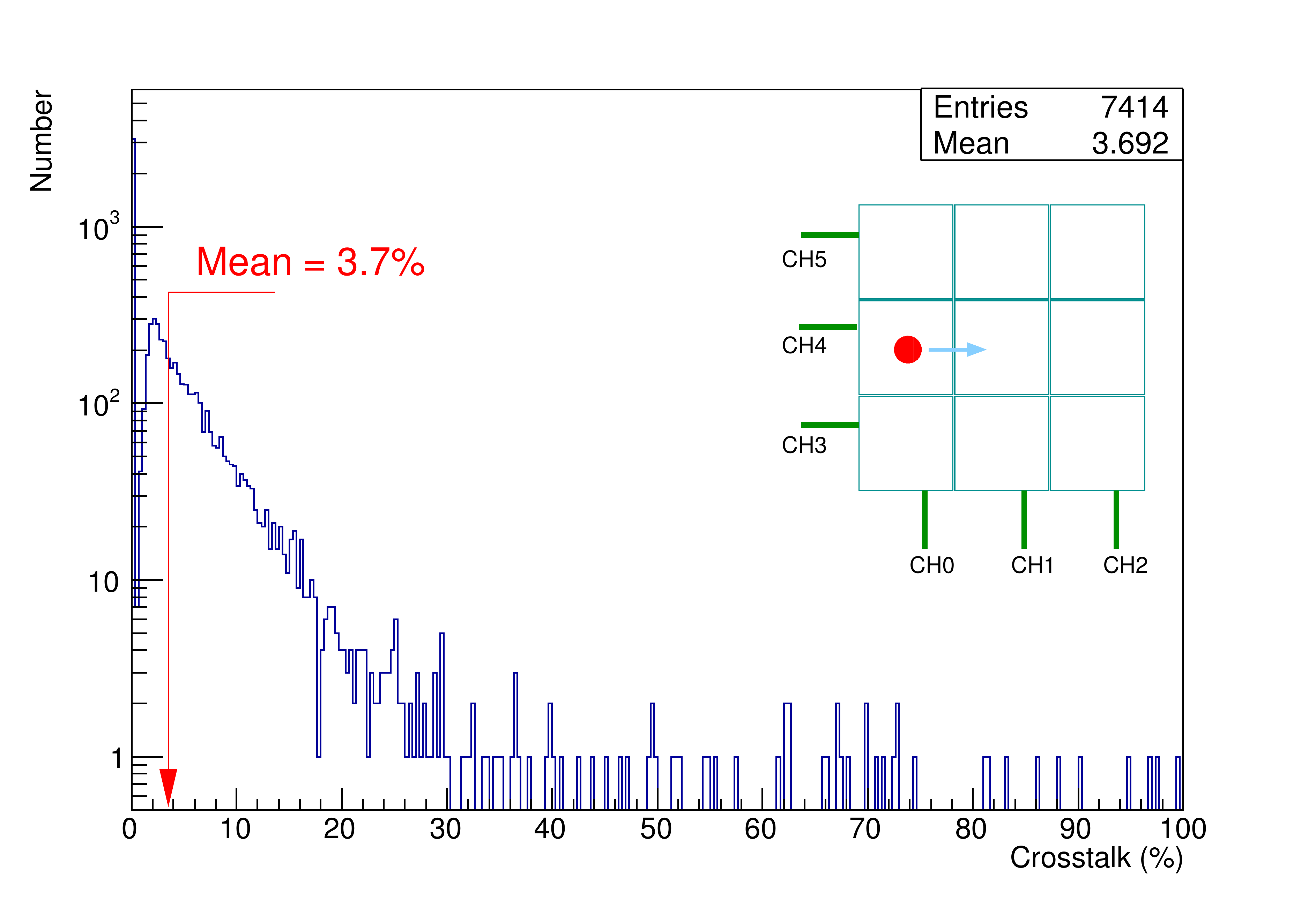}
\caption{ Crosstalk from the cube ch0/ch4  into the cube ch1/ch4. Beam hits the central part of the  ch0/ch4  cube.}
\label{fig:cross_spec} 
\end{centering}
\end{figure}

 Fig.~\ref{fig:cross_spec} shows the crosstalk distribution when the light from the cube CH0/CH4 leaks into the cube CH1/CH4 (see Fig.~\ref{fig:readout} for the channel labelling). The crosstalk was calculated as the ratio $L.Y._{CH1}/L.Y._{CH0}$. The crosstalk average value is 3.7\%, while the average of $L.Y._{CH0}$ is 41~p.e.  
%========================
%======================== 
The crosstalk with values higher than 30\% can be explained by shower events.

\begin{figure}[h]
\begin{centering}
\includegraphics[width=5cm]{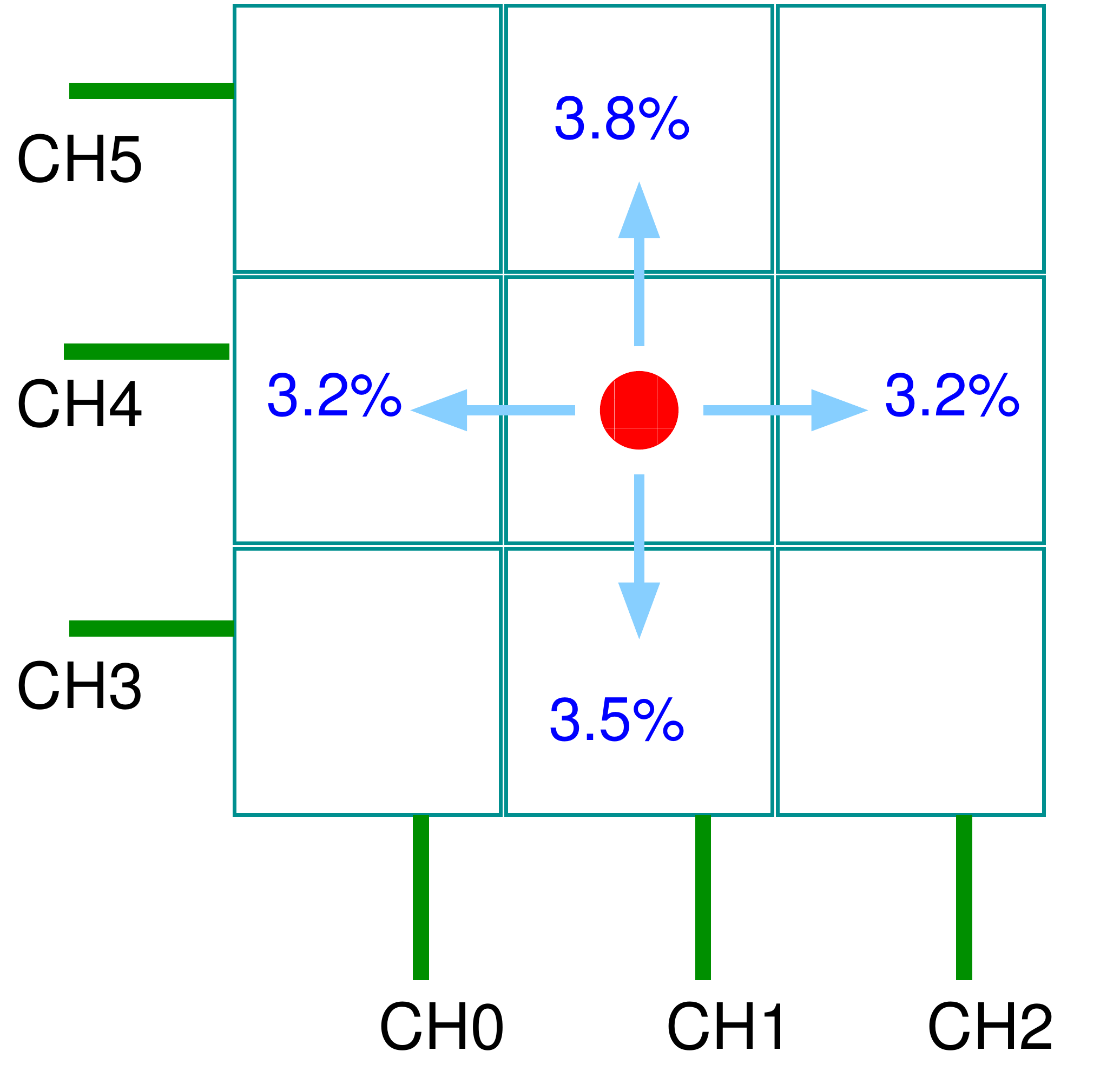}
\caption{ Crosstalk from the central cube in four directions.}
\label{fig:cross4} 
\end{centering}
\end{figure}

The crosstalk in four directions is shown in Fig.~\ref{fig:cross4}, when the beam hits the central cube CH1/CH4 in a $3\times 3$ array.  The average crosstalk is 3.4\% per side. The average of the total crosstalk into all 4 sides on an event-to-event basis was measured to be 13.7\%. From this we can conclude that ~20\% of the detected scintillation light escapes the fired cube into adjacent cubes through the cube reflective walls.
%========================
%======================== 

\subsection{Time resolution}

We have applied the constant fraction method to obtain the timing mark of the signal waveform. The preamplifiers extend the signal front to 7~ns  (measured between 0.1--0.9 fractions of the amplitude), so that we have up to 40 digitizer sample points spaced at 200~ps at the front. The following procedure was used for each waveform to obtain the timing.  First, the baseline was determined by fitting the first sample points before a signal with a horizontal line. Then a maximum amplitude of the waveform was measured.  We have found that a  fraction of 10\% of the maximum amplitude at the signal front provides the best timing, so the next step was to determine which point on the time axis intercepts the amplitude level of  10\%  of the maximum. 

\begin{figure}[h]
\begin{centering}
\includegraphics[width=8cm]{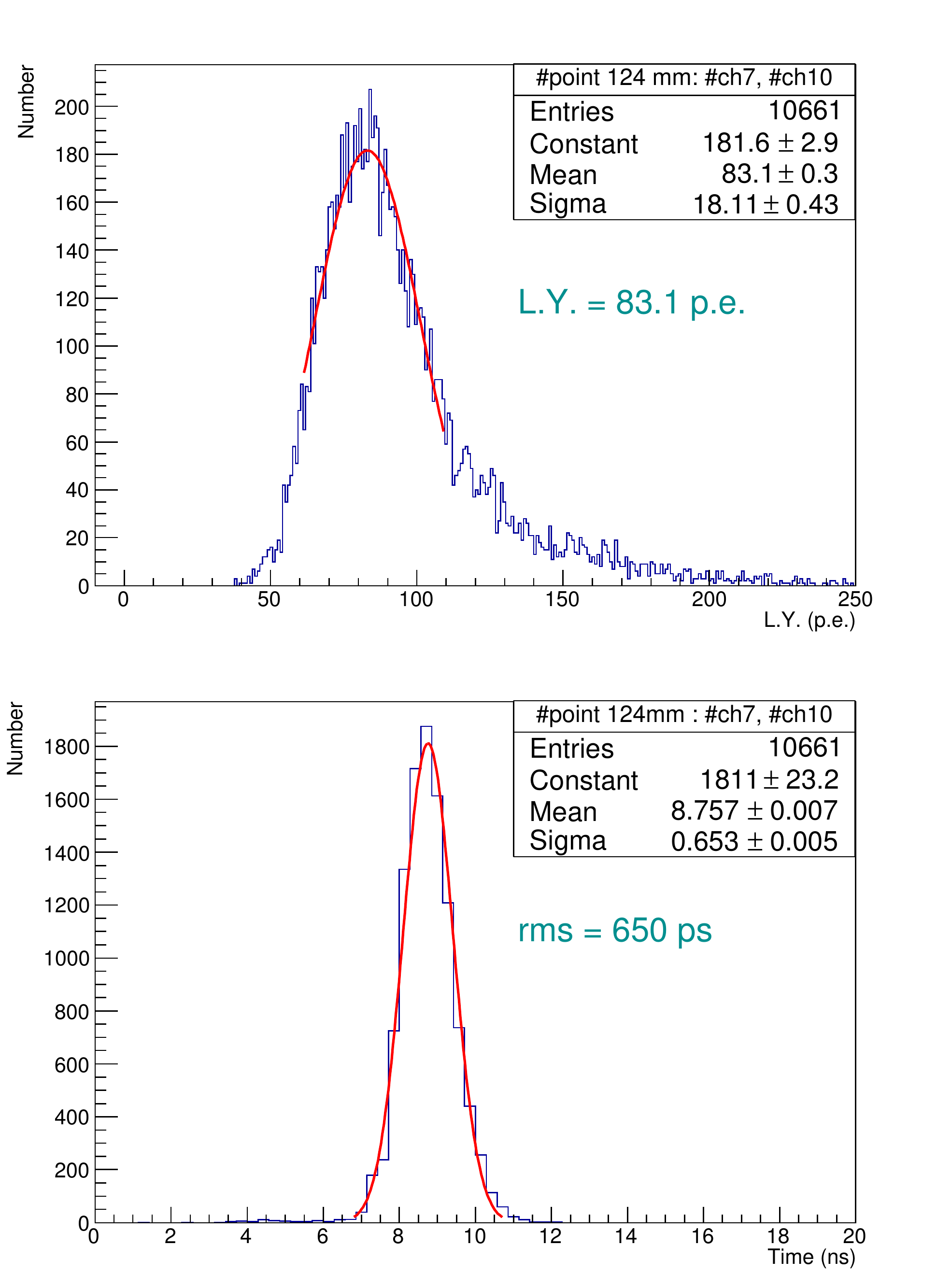}
\caption{ Charge and time spectra for a single cube. Charge signal is a sum from two fibers, the time is an average time between two fibers. }
\label{fig:spectra} 
\end{centering}
\end{figure}

Two options were exploited to get this time mark. First one is to find the first closest sample point at time axis which corresponds to the calculated value of 0.1 of the maximum. A more precise but more complicated option requires extrapolation between two closest sample points on the time axis which correspond to the amplitudes below and above the level of 0.1 of the maximum. The second method provides slightly better  timing at the digitizer sampling rate of 5~GHz, and the advantage becomes more significant at the slower rates.
The digitizer itself generates a time jitter correlated in all the channels. This source of time fluctuation was removed by operating a  subtraction between the measured  channel and the trigger counter on an event-by-event basis. For a single fiber we calculated $T_{fiber}=T_{ch}-T_{tr}$,  the timing of a cube was obtained as the average time between the two  readout fibers.  Fig.~\ref{fig:spectra} shows the time and charge distributions for one of the cubes. 
%========================
%======================== 

Typical time resolution ($\sigma_t$) for a single fiber was around 0.95~ns.  A cube with two readout fibers gives $\sigma_t$=0.65--0.71~ns. Two cubes combined produced typical $\sigma_t$=0.52~ns for the first method of the time mark calculation, and  $\sigma_t$=0.48~ns for the second method.

\section{Summary}

The superFGD prototype with 3D fiber readout was  assembled of 125 plastic scintillator cubes of $1\times 1\times 1~cm^3$ size  and successfully tested in a charge particles beam at CERN. Hamamatsu MPPCs  viewed 1.3~m long 1~mm diameter WLS fibers Kuraray Y11 at a single end. A 5~GHz digitizer was used to read out 12 fibers out of 75 mounted ones to measure the parameters of the prototype.

Typical light signal in a cube was about  40~p.e. per a single fiber. About 20\% of the detected scintillating light escapes a fired cube into adjacent cubes through the cube reflective walls. Leakage from a fired cube  into a neighboring cube through a side was found to be between 3.5--4.0\%. Typical time resolution  for a single fiber was around 0.95~ns.  A cube with two readout fibers gives $\sigma_t$=0.65--0.71~ns and two cubes with four fibers produced typical $\sigma_t$=0.48~ns.

We plan to expose a  superFGD prototype~\cite{sfgd_neutrino2018} made of 9200 cubes  in a CERN test beam in summer 2018, where tracking capabilities of the detector will be investigated.

%_________________________________________________________________________

\section*{Acknowledgments}

This work was in part supported   by the RFBR/JSPS  grant \# 17-52-50038 and by the RFBR grant \# 18-32-00070.
We are grateful to all  collaboration members for the valuable cooperation and detailed discussions regarding the superFGD prototype. 

%===============================

\end{document}